\documentclass[aps,prd,preprintnumbers,amsmath,showkeys,twocolumn,amssymb,floatfixng,nofootinbib,a4paper]{revtex4-1}
\usepackage{graphicx,slashed,booktabs,xcolor,multirow,float,amsfonts,bbold,mathtools,sidecap,tikz,bm,ulem,enumitem}
\usepackage[colorlinks=true,linktoc=page,linkcolor=purple,citecolor=teal,urlcolor=magenta]{hyperref}
\pdfoutput=1
\allowdisplaybreaks

\begin{document}
\preprint{IP/BBSR/2021-08}

\definecolor{lime}{HTML}{A6CE39}
\DeclareRobustCommand{\orcidicon}{\hspace{-2.1mm}
\begin{tikzpicture}
\draw[lime,fill=lime] (0,0.0) circle [radius=0.13] node[white] {{\fontfamily{qag}\selectfont \tiny \,ID}}; \draw[white, fill=white] (-0.0525,0.095) circle [radius=0.007]; 
\end{tikzpicture} \hspace{-3.7mm} }
\foreach \x in {A, ..., Z}{\expandafter\xdef\csname orcid\x\endcsname{\noexpand\href{https://orcid.org/\csname orcidauthor\x\endcsname} {\noexpand\orcidicon}}}
\newcommand{\orcidauthorA}{0000-0001-5643-2652}

\title{Type-III see-saw: Search for triplet fermions in final states with multiple leptons and fat-jets at 13 TeV LHC}
\author{Saiyad Ashanujjaman\orcidA{}}
\email{saiyad.a@iopb.res.in}
\author{Kirtiman Ghosh}
\email{kirti.gh@gmail.com}
\affiliation{Institute of Physics, Bhubaneswar, Sachivalaya Marg, Sainik School, Bhubaneswar 751005, India}%                                                           
\affiliation{Homi Bhabha National Institute, Training School Complex, Anushakti Nagar, Mumbai 400094, India}%                                                                                           
%\date{\today}
                                                                                               
\begin{abstract}
\noindent {The type-III see-saw model holding out a riveting rationale for the minuscule neutrino masses caters for a well-to-do phenomenology at the Large Hadron Collider (LHC). Several searches targetting the triplet fermions have been performed at the LHC. Not only are the signals for the leptonic final states considered in these searches suppressed by the branching fractions of the Standard Model (SM) bosons, but they are also beset with considerably large SM backgrounds. Thus, these searches are deemed not to be sensitive enough in probing the triplet fermions much heavier than 1 TeV. To this end, we perform a search for the triplet fermions in final states with multiple leptons and fat-jets that are cleaner than the usual LHC searches and allow kinematic reconstruction of the triplets. After performing a systematic and comprehensive analysis with seven distinct final states, we project the required luminosities for both $3\sigma$ and $5\sigma$ discoveries of the triplet fermions as a function of their mass. The triplet fermions with mass as large as 1265 (1380) and 1480 (1600) GeV could be discovered with $5\sigma$ ($3\sigma$) significance at 500 and 3000 fb$^{-1}$ luminosities, respectively.}
\end{abstract}
\maketitle

\section{\label{sec:intro}Introduction}
Despite being surpassingly successful, the Standard Model (SM) falls short of proffering a natural and well-founded rationale for the observed sub-eV neutrino masses. Diversely, the type-III see-saw model, one of the three UV completions of the so-called Weinberg operator \cite{Weinberg:1979sa} at the tree-level \cite{Ma:1998dn}, based on the annexation of the SM by weak gauge triplet of fermion field \cite{Foot:1988aq} seems to explain the minuteness of neutrino masses readily. Though the see-saw anchors are naturally motivated to have very high scale masses, if balanced with appropriate Yukawa couplings, nothing precludes them from having mass at TeV scale, and thus a well-to-do phenomenology at the Large Hadron Collider (LHC). Phenomenological outcome of this model has been studied all-encompassingly in the literature \cite{Bajc:2006ia,FileviezPerez:2007bcw,FileviezPerez:2007yji,delAguila:2008hw,delAguila:2008cj,Franceschini:2008pz,Arhrib:2009mz,Bandyopadhyay:2011aa,Biggio:2011ja,Goswami:2017jqs,Cai:2017mow,Das:2020gnt,Jana:2020qzn,Das:2020uer,Ashanujjaman:2021jhi,Sen:2021fha}. At the LHC, a number of collider searches looking for the triplet fermions have been carried out \cite{ATLAS:2015fma,CMS:2017ybg,ATLAS:2018ghc,CMS:2019lwf,ATLAS:2020wop}, and the observations being consistent with the SM expectations, these searches have derived exclusion limits at 95\% confidence level (CL) on the mass of the triplet fermions. The most stringent limit of 880 GeV has been put by the recent CMS multi-lepton search \cite{CMS:2019lwf} assuming only one generation of them in the flavour-democratic scenario. This limit extends to 1110 GeV when three mass-degenerate triplets\footnote{Three generations of triplet fermions are required to generate three non-vanishing neutrino masses. Note that one (two) generation(s) of them would not suffice to explain more than one (two) non-zero neutrino masses.} are assumed \cite{Ashanujjaman:2021jhi}.

Not only are the signals for the leptonic final states considered in the abovementioned LHC searches \cite{ATLAS:2015fma,CMS:2017ybg,ATLAS:2018ghc,CMS:2019lwf,ATLAS:2020wop} suppressed by the leptonic branching fractions of the SM bosons ($W,Z,h$), but they are also beset with considerably large SM backgrounds. Consequently, these searches are deemed not to be sensitive enough in probing the triplet fermions much heavier than 1 TeV. Moreover, these searches are not designed to reconstruct the triplet mass. This behoves us to look for complementary final states, such as multiple leptons and fat-jets resulting from the hadronically decaying SM bosons, which are not only clean and thus sensitive enough in probing the TeV scale triplets but also allow kinematic reconstruction of them. To this end, we perform a systematic and comprehensive search for the triplet fermions in final states with multiple leptons and fat-jets.

The rest of this work is structured as follows. In Section~\ref{sec:model}, we briefly discuss the productions and decays of the triplet fermions. We perform a systematic and comprehensive collider analysis for the triplet fermions in final states with multiple leptons and fat-jets at the 13 TeV LHC in Section~\ref{sec:collider} followed by a summary in Section~\ref{sec:conclusion}.

\section{\label{sec:model} The triplet fermions} 
The type-III see-saw model employs, in addition to the SM field content, three generations of right-handed $SU(2)_L$ triplet fermions with vanishing hypercharge
\begin{equation*}
\Sigma = \left( \begin{array}{cc} \Sigma^0_R/\sqrt{2} & \Sigma^+_R \\ \Sigma^-_R & -\Sigma^0_R/\sqrt{2} \end{array} \right);
\end{equation*}
and the relevant triplet Yukawa and mass terms are given by
\begin{equation*}
-{\cal L} \supset \frac{1}{2} \left( M_\Sigma \right)_{ij} {\rm Tr}\left( ~\overline{\tilde{\Sigma}_i} \Sigma_j \right) + \sqrt{2}~ \left( Y_\Sigma \right)_{i \alpha}~\tilde{H}^\dagger \overline{\Sigma_i} L_\alpha ~ + \rm{h.c.},
\end{equation*}
where $i,j$ and $\alpha$ are the generation indices which run over 1,2 and 3, $M_\Sigma$ is the Majorana mass matrix for $\Sigma$, $\tilde{\Sigma}$ denotes charge-conjugation of $\Sigma$, {\it i.e.} $\tilde{\Sigma}=C\overline{\Sigma}^T$ with $C$ being the charge-conjugation matrix, $Y_\Sigma$ is the Yukawa matrix, $L=\left(\nu_L, \ell_L \right)^T$ is the left-handed SM lepton doublet, and $H$ is the SM Higgs doublet. For simplicity, we assume $M_\Sigma = m_\Sigma \mathbb{1}_{3\times 3}$, {\it i.e.} all three generations of triplets are mass-degenerate. The above Lagrangian leads to the so-called {\it see-saw formula} for the light neutrinos: $m_\nu \approx -\frac{v^2}{2} Y_\Sigma^T M_\Sigma^{-1} Y_\Sigma $. Fo a given $M_\Sigma$, the well-known {\it Casas-Ibarra parametrisation} \cite{Casas:2001sr,Ibarra:2003up} enables one to find the most general texture of $Y_\Sigma$ to reproduce the low-energy neutrino oscillation parameters \cite{Esteban:2018azc} which we avert to discuss in this work, see \cite{Jana:2020qzn,Das:2020uer,Ashanujjaman:2021jhi}.

%Production 
The triplet fermions are pair produced copiously at the LHC by quark-antiquark annihilation via $s$-channel $\gamma/Z$ and $W^{±}$ exchanges $$ q\bar{q^\prime} \to W^* \to \Sigma^\pm \Sigma^0, \quad {\rm and} \quad q\bar{q} \to \gamma^*/Z^* \to \Sigma^+ \Sigma^-.$$ They are also pair produced via photon-photon fusion processes:\footnote{They are also pair produced via vector-boson fusion processes with one or two associated forward jets at the LHC. However, such processes are rather sub-dominant and can be neglected.} $ \gamma \gamma \to \Sigma^+ \Sigma^-.$ Their production at the LHC has been widely studied in the literature \cite{Bajc:2006ia,FileviezPerez:2007bcw,FileviezPerez:2007yji,delAguila:2008hw,delAguila:2008cj,Franceschini:2008pz,Arhrib:2009mz,Bandyopadhyay:2011aa,Biggio:2011ja,Goswami:2017jqs,Cai:2017mow,Das:2020gnt,Jana:2020qzn,Das:2020uer,Ashanujjaman:2021jhi,Sen:2021fha}. In our analysis, we evaluate the leading order (LO) production cross-sections using the UFO modules generated from SARAH \cite{Staub:2013tta,Staub:2015kfa} in MadGraph \cite{Alwall:2011uj,Alwall:2014hca} with the {\it LUXqed17-plus-PDF4LHC15-nnlo-100} parton distribution function \cite{Manohar:2016nzj,Manohar:2017eqh,Butterworth:2015oua}. Following Ref.~\cite{Ruiz:2015zca}, we naively scale the LO cross-section by an overall $K$-factor of 1.30.\footnote{Ref.~\cite{Ruiz:2015zca} has reported the leading QCD corrections to the production of triplet fermions at hadron colliders. The resulting next-to-leading order (NLO) $K$-factors varies from 1.24 to 1.37 over a triplet mass range of 1 TeV to 2 TeV.} The total pair production cross-sections varies from 585 fb to 0.02 fb for 400 GeV to 2 TeV triplet mass.
%\begin{figure}[htb!]
%\centering
%\includegraphics[width=0.85\columnwidth]{xsec}
%\caption{Total triplet production cross section.}
%\label{fig:cs}
%\end{figure}
 
%Decay  
After being produced, the triplet fermions undergo two types of decays --- $(i)$ the {\it heavy state transitions} due to the radiative mass-splitting of $\Delta m=m_{\Sigma^+}-m_{\Sigma^0} \sim 167$ MeV \cite{Cirelli:2005uq} , and $(ii)$ the {\it two-body decays} into a SM boson ($W,Z,h$) and a lepton ($\ell,\nu$). For the Yukawa couplings $\left(Y_\Sigma \right)_{i\alpha}$ not smaller than $\mathcal{O}(10^{-8})$, the later decay modes subjugate the former ones ensuring their prompt decays into two SM particles \cite{Jana:2020qzn,Das:2020uer,Ashanujjaman:2021jhi}. The relevant partial decay widths are given by
\begin{equation*}
\Gamma(\Sigma^{0(\pm)}_i\to XY^{(\prime)}_\alpha)  = \kappa \frac{m_\Sigma}{64\pi} |(Y_\Sigma)_{i\alpha}|^2 f\left(\frac{m_X}{m_\Sigma}\right),
\end{equation*}
where $XY_\alpha \ni h \nu_\alpha,Z \nu_\alpha,W^\pm \ell^\mp_\alpha$, $XY^\prime_\alpha \ni h \ell^\pm_\alpha,Z \ell^\pm_\alpha,W^\pm  \nu_\alpha$, $f(r)=(1-r^2)^2(1+\kappa^\prime r^2),\kappa(\kappa^\prime)=1,1,2(0,2,2)$ for $X=h,Z,W^\pm$, respectively. The branching fractions for the $h,Z$ and $W$ decay modes are, respectively, 25\%, 25\% and 50\%. For simplicity, all three generations of triplets are assumed to be mass-degenerate.\footnote{While, {\it a priori}, nothing precludes the mass from varying across the generations, including such a possibility only introduces additional parameters (and thus complications) without adding anything qualitatively different to the discussion.} Further, we assume identical branching fractions across all lepton flavours ({\it flavour-democratic scenario}).\footnote{We accentuate that the flavour democratic scenario does not necessarily be an {\it ad-hoc} assumption; rather, it could be a consequence of degenerate mass spectrum for both the light and heavy neutrinos or could be a repercussion of the information lost in integrating the heavy leptons out \cite{Ashanujjaman:2021jhi}. Also, this assumption facilitates an easy comparison with existing search paradigms adopted by the ATLAS/CMS collaborations. However, relaxation of this assumption may significantly alter the phenomenology, and thus the exclusion limit, see \cite{Ashanujjaman:2021jhi}.} This is readily achieved by choosing the triplet Yukawa couplings that are compatible with the existing constraints \cite{Abada:2008ea,Goswami:2018jar,Biggio:2019eeo} as well as the neutrino oscillation data \cite{Esteban:2018azc}, see \cite{Jana:2020qzn,Das:2020uer,Ashanujjaman:2021jhi}.

\section{\label{sec:collider}Collider phenomenology}
At the LHC, the triplet fermions are pair produced aplenty followed by their prompt decays to a SM lepton ($\ell,\nu$) and a boson ($W,Z,h$) leading to multifarious final state signatures. The characteristic signature of TeV scale triplets\footnote{The recent CMS multi-lepton search \cite{CMS:2019lwf} has already excluded triplet fermions with masses below 880 GeV assuming only one generation of them in the flavour-democratic scenario. This limit extends to 1110 GeV when three mass-degenerate triplets are considered \cite{Ashanujjaman:2021jhi}.} would be vastly boosted final sate lepton and boson which are likely to be largely separated in the azimuth plane. The jets stemming from the sufficiently highly boosted boson would be collimated. The hadronically decaying bosons, thus, are more likely to manifest as a single fat-jet rather than two resolved jets. In what follows, we perform a systematic and comprehensive collider study for the triplet fermions in the final states with multiple leptons and fat-jets. The processes contributing to the final state signatures of our interest are summarised in the following:
 
\noindent $(i)$ {\it One lepton and fat-jets:} Both pair and associated production of charged triplets contribute to this final state:
\begin{enumerate}[label=(\alph*)]
\item $\Sigma^\pm \Sigma^\mp \to \ell^\pm Z/h \nu W^\mp \to \ell^\pm \nu J_{Z/h} J_W$,
\item $\Sigma^+ \Sigma^- \to \nu W^+ \nu W^- \to \ell^\pm \nu \nu \nu J_W$,
\item $\Sigma^\pm \Sigma^0 \to \ell^\pm Z/h \nu Z/h \to \ell^\pm \nu J_{Z/h} J_{Z/h}$,
\item $\Sigma^\pm \Sigma^0 \to \nu W^\pm \ell^\pm W^\mp \to \ell^\pm \nu J_{W} J_{W}$,
\end{enumerate}
where $J_{W/Z/h}$ denotes a fat-jet emanating from $W/Z/h$. For process (b), neutrinos result from both triplet fermions making kinematic reconstruction of the latter challenging. For all other processes, one of the triplets decays to $\ell^\pm$ and $W/Z/h$, where the latter manifests as $J_{W/Z/h}$. The triplets being in the TeV regime, $\ell^\pm$ and $J_{W/Z/h}$ are deemed to be largely separated in the azimuth plane. Therefore, $m_\Sigma$ can be reconstructed from the invariant mass distribution of the lepton and the farthest $J_{ZW//h}$ in the azimuth plane.

\noindent $(ii)$ {\it Same-sign dilepton and fat-jets:} Only the following process contributes to this final state:
$$\Sigma^\pm \Sigma^0 \to \ell^\pm Z/h \ell^\pm W^\mp \to \ell^\pm \ell^\pm J_{Z/h} J_W.$$
The invariant mass distribution of $J_{W/Z/h}$ and the farthest lepton in the azimuth plane would peak at $m_\Sigma$.

\noindent $(iii)$ {\it Opposite-sign dilepton and fat-jets:} The processes contributing to this final states are as follows:
\begin{enumerate}[label=(\alph*)]
\item $\Sigma^+ \Sigma^- \to \ell^+ Z/h \ell^- Z/h \to \ell^+ \ell^- J_{Z/h} J_{Z/h}$,
\item $\Sigma^\pm \Sigma^\mp \to \ell^\pm Z/h \nu W^\mp \to \ell^\pm \ell^\mp \nu \nu J_{Z/h}$,
\item $\Sigma^\pm \Sigma^0 \to \ell^\pm Z/h \ell^\mp W^\pm \to \ell^\pm \ell^\mp J_{Z/h} J_W$.
\end{enumerate}
Similar to the above channel, the invariant mass distribution of $J_{W/Z/h}$ and the farthest lepton in the azimuth plane would yield $m_\Sigma$.

\noindent $(iv)$ {\it Trilepton without fat-jet:} Both pair and associated production of charged triplets contribute to this final state:
\begin{enumerate}[label=(\alph*)]
\item $\Sigma^\pm \Sigma^0 \to \nu W^\pm \ell^{\pm(\mp)} W^{\mp(\pm)} \to \ell^\pm \ell^\mp \ell^\pm \nu \nu$,
\item $\Sigma^\pm \Sigma^0 \to \nu W^\pm \nu Z/h \to \ell^\pm \ell^+ \ell^- \nu \nu$,
\end{enumerate}
Here, both the triplets lead to a neutrino in the final state, making their kinematic reconstruction challenging.

\noindent $(v)$ {\it Trilepton and one fat-jet:} Both pair and associated production of charged triplets contribute to this final state:
\begin{enumerate}[label=(\alph*)]
\item $\Sigma^\pm \Sigma^\mp \to \ell^\pm Z/h \nu W^\mp \to \ell^\pm \ell^+ \ell^- \nu J_W$,
\item $\Sigma^\pm \Sigma^0 \to \ell^\pm Z/h \nu Z/h \to \ell^\pm \ell^+ \ell^- \nu J_{Z/h}$,
\item $\Sigma^\pm \Sigma^0 \to \ell^\pm Z/h \ell^{\pm(\mp)} W^{\mp(\pm)} \to \ell^\pm \ell^\pm \ell^\mp \nu J_{Z/h}$,
\end{enumerate}
For processes (a) and (b), $\Sigma^\pm$ decays to $\ell^\pm$ and $Z/h$ which further decays to $\ell^+ \ell^-$. For process (c), $\Sigma^\pm$ decays to $\ell^\pm$ and $J_{Z/h}$. Therefore, for the events with an on-$Z$ lepton pair,\footnote{An opposite-sign same flavour lepton pair with an invariant mass within the $Z$-boson mass window, {\it i.e.} $m_Z\pm 15$ GeV is termed as ``on-$Z$ lepton pair".} the peak of the invariant mass distribution of the trilepton system is expected to yield $m_\Sigma$, whereas, for those with no on-$Z$ lepton pair, $m_\Sigma$ can be reconstructed from the invariant mass distribution of $J_{Z/h}$ and the farthest lepton in the azimuth plane.

\noindent $(vi)$ {\it Tetralepton with/without one fat-jet:} Both pair and associated production of charged triplets contribute to this final state:
\begin{enumerate}[label=(\alph*)]
\item $\Sigma^+ \Sigma^- \to \ell^+ Z/h \ell^- Z/h \to \ell^+ \ell^+ \ell^- \ell^- J_{Z/h}$.
\item $\Sigma^\pm \Sigma^\mp \to \ell^\pm Z/h \nu W^\mp \to \ell^\pm \ell^+ \ell^- \ell^\mp \nu \nu$.
\item $\Sigma^\pm \Sigma^0 \to \ell^\pm Z/h \ell^{\pm(\mp)} W^{\mp(\pm)} \to \ell^\pm \ell^+ \ell^- \ell^{\pm(\mp)} J_W$.
\end{enumerate}
All the events necessarily contain an on-$Z$ lepton pair. The invariant mass distribution of the on-$Z$ lepton pair and the farthest lepton between the ones which are not part of the on-$Z$ lepton pair would peak at $m_\Sigma$.

We next give a brief description of reconstruction and selection of various physics objects, classification of events into several mutually exclusive analysis channels and event selection.

\subsection{Object reconstruction and selection} We use Delphes \cite{deFavereau:2013fsa} to reconstruct different objects, namely photons, electrons, muons and jets. Constituents of the {\it would-be} fat-jets are clustered using the {\it anti-k$_T$ algorithm} \cite{Cacciari:2008gp} with the winner-take-all axis with the characteristic fat-jet radius $R=0.8$ as implemented in FastJet \cite{Cacciari:2011ma}. We use the {\it jet pruning} algorithm \cite{Ellis:2009su,Ellis:2009me}, setting the parameters to their default values, {\it viz.} $z_{cut} = 0.1$ and $R_{cut} = 0.5$ \cite{Ellis:2009su}, to remove the softer and wide-angle QCD emissions from the fat-jets. Further, an inclusive jet shape termed as {\it $N$-subjettiness}, $\tau_N$, \cite{Thaler:2010tr,Thaler:2011gf}  is used to unfold the multi-prong nature of the fat-jets. We choose the {\it one-pass $k_T$-axes} for the minimization procedure, and take the thurst parameter $\beta=1$. The reconstructed leptons (electrons and muons) and fat-jets are required to have transverse momentum $p_T > 10$ and 30 GeV, respectively, and they must lie within the tracking system acceptance, pseudorapidity $|\eta|<2.5$. The relative isolation\footnote{The relative isolation of a lepton is defined as the scalar $p_T$ sum, normalized to the lepton $p_T$, of photons and hadrons within a cone of $\Delta R=0.5$ around the lepton.} is required to be smaller than 12\%(15\%) for electrons (muons). Such stringent lepton isolation requirements significantly suppress the reducible backgrounds. The selected leptons and jets are ordered by decreasing $p_T$ ($\ell_0,\ell_1,...$ and $J_0,J_1,...$). Often, some of the jets are misidentified as leptons. Following the Ref.~\cite{ATLAS:2016iqc}, we take the probability of 0.1--0.3\% for a jet to be misidentified as a lepton. Furthermore, bremsstrahlung interactions of electrons with the inner detector material often lead to their charge misidentification. We adopt the charge misidentification probability from Ref.~\cite{ATLAS:2017xqs}: $P(p_T,\eta)=\sigma(p_T) \times f(\eta)$, where $\sigma(p_T)$ and $f(\eta)$ ranges from 0.02 to 0.1 and 0.03 to 1, respectively. Finally, the missing transverse momentum vector $\vec p_T^{\rm ~miss}$ (with magnitude $p_T^{\rm miss}$) is estimated using all the reconstructed particle-flow objects in an event.

\subsection{SM Backgrounds} 
We next briefly mention the SM processes which could resemble the abovementioned final states. The relevant processes can be classified into two classes --- reducible and irreducible backgrounds. The SM processes giving rise to one or more prompt leptons, such as Drell-Yan processes, $VV$ ($V=W,Z/\gamma^*$), $VVV$, $VVVV$, $t\bar{t}$, $t\bar{t}V$, single top production ($tb,tW,tj$) and Higgsstrahlung ($Vh, VVh, t\bar{t}h$) processes, constitute the irreducible backgrounds in this analysis. On the contrary, the reducible backgrounds are from the SM processes like $Z/\gamma^*+$jets and $t\bar{t}+$jets, where a jet is misidentified as lepton or additional leptons originate from heavy quark decays. The reducible backgrounds are significantly suppressed by applying stringent lepton isolation requirements. All the background samples are generated in association with up to two jets using MadGraph \cite{Alwall:2011uj,Alwall:2014hca} at the LO using the 5 flavour scheme followed by MLM matching in PYTHIA \cite{Sjostrand:2014zea}. The background samples are normalised at least to the NLO prediction \cite{Campbell:1999ah,Ciccolini:2003jy,Brein:2003wg,Catani:2007vq,Campanario:2008yg,Balossini:2009sa,Bredenstein:2009aj,Catani:2009sm,Kidonakis:2010ux,Campbell:2011bn,Brein:2011vx,Bevilacqua:2012em,Garzelli:2012bn,Brein:2012ne,Altenkamp:2012sx,Nhung:2013jta,Kidonakis:2013zqa,Denner:2014cla,Harlander:2014wda,Kidonakis:2015nna,Muselli:2015kba,Shen:2015cwj,Frederix:2017wme}.

%\subsection{\label{sec:channels} Analysis channels} 
%After object selection, we make our identification of the $W/Z/h$-fat-jets, denoted as $J_{W/Z/h}$, satisfying the conditions:\footnote{The {\it N-subjettiness}, $\tau_N$, is a good measure of the number of subjets a jet is presumably composed of. The ratio $\tau_N/\tau_{N-1}$ is, thus, an useful discriminant between the $N$- and $(N-1)$-prong jets. Being two-prong in nature, the jets emanating from the sufficiently highly boosted $W/Z/h$-bosons tend to have lower $\tau_{21}=\frac{\tau_2}{\tau_1}$ as compared to the overwhelming one-prong QCD or top jets. The cut on $\tau_{21}$ significantly suppresses the backgroud contributions from the QCD or top-jets.}
%\[ m(J) \in [70,140] {\rm ~GeV,~} p_T(J) > 200 {\rm ~GeV ~and~} \tau_{21} < 0.7. \]
%Based on the number of leptons and $J_{W/Z/h}$, the events are categorised into the following seven distinct final states (analysis channels) while the events not falling under these channels are thrown away: 
%\begin{enumerate}[label=(\roman*)]
%\item \textbf{\textit{1L-1J:}} one lepton and one $J_{W/Z/h}$,
%\item \textbf{\textit{1L-2J:}} one lepton and two $J_{W/Z/h}$,
%\item \textbf{\textit{SSD-1J:}} same-sign dilepton and $\geq 1$ $J_{W/Z/h}$,
%\item \textbf{\textit{OSD-1J:}} opposite-sign dilepton $\geq 1$ $J_{W/Z/h}$,
%\item \textbf{\textit{3L-0J:}} trilepton without any $J_{W/Z/h}$,
%\item \textbf{\textit{3L-1J:}} trilepton and one $J_{W/Z/h}$,
%\item \textbf{\textit{4L:}} tetralepton and any number of $J_{W/Z/h}$.
%\end{enumerate}

\subsection{\label{sec:sel} Event selection} 
After object selection, we make our identification of the $W/Z/h$-fat-jets, denoted as $J_{W/Z/h}$, satisfying the conditions:\footnote{The {\it N-subjettiness}, $\tau_N$, is a good measure of the number of subjets a jet is presumably composed of. The ratio $\tau_N/\tau_{N-1}$ is, thus, an useful discriminant between the $N$- and $(N-1)$-prong jets. Being two-prong in nature, the jets emanating from the sufficiently highly boosted $W/Z/h$-bosons tend to have lower $\tau_{21}=\frac{\tau_2}{\tau_1}$ as compared to the overwhelming one-prong QCD or top jets. The cut on $\tau_{21}$ significantly suppresses the background contributions from the QCD or top-jets.}
\[
m(J) \in [70,140] {\rm ~GeV,~} p_T(J) > 200 {\rm ~GeV ~and~} \tau_{21} < 0.7.
\]
Based on the number of leptons and $J_{W/Z/h}$ ($n_\ell$ and $n_{J_{W/Z/h}}$) and the absolute value of the sum of charges of the leptons $\left(|\sum Q_\ell|\right)$, the events are categorised into seven distinct final states --- {\it 1L-1J}, {\it 1L-2J}, {\it SSD-1J}, {\it OSD-1J}, {\it 3L-0J}, {\it 3L-1J} and {\it 4L} (see Table~\ref{preselection}), while the events not falling under these channels are thrown away. As mentioned earlier, the leptons and bosons coming from the decays of the TeV sacle triplets would carry sufficiently high momenta that stronger cuts on the leptons' $p_T$ and/or $p_T^{\rm miss}$ would be utile in curtailing the SM background while keeping the signal all but unharmed. The event selection proceeds in two steps: the preselection and the channel-specific event selection. The preselection requirements are primarily based on $p_T(\ell_0)$, $p_T(\ell_1)$ and $p_T^{\rm miss}$, see Table~\ref{preselection}. %In the {\it 1L-1J} and {\it 1L-2J} channels, the events with $p_T(\ell_0) > 200$ GeV and $p_T^{\rm miss} > 100$ GeV are considered for further analysis. In the {\it SSD-1J} and {\it OSD-2J} channels, the leading and subleading leptons are required to have $p_T > 400$ and $100$ GeV, respectively. Similarly, the leading two leptons in the {\it 3L-0J} and {\it 3L-1J} channels are required to have $p_T > 300$ and $100$ GeV, respectively. Further, the {\it 3L-0J} events with $p_T^{\rm miss} < 100$ GeV are rejected. In the {\it 4L} channel, the events with $p_T(\ell_0) > 400$ GeV and $p_T(\ell_1) > 200$ GeV are considered further. Furthermore, 
Events containing an opposite-sign same-flavour (OSSF) lepton pair with invariant mass below 12 GeV are thrown away in all the channels to reduce background contributions from low-mass resonances. Furthermore, to suppress the overwhelming background contributions from the Drell-Yan processes in the {\it SSD-1J} and {\it OSD-2J} channels, events containing an opposite-sign same-flavour lepton pair with an invariant mass within the nominal $Z$-boson mass window, {\it i.e.} $m_Z \pm 15$ GeV are vetoed.\footnote{Note that, in order to reduce the contributions originating from electron charge misidentification, the Z-veto is also applied to the {\it SSD-1J} events.}

\begin{table}[htb!]
\centering
\scalebox{0.86}{
\begin{tabular}{|c|c|c|c|c|c|c|c|}
\hline \hline 
Selection criteria & {\it 1L-1J} & {\it 1L-2J} & {\it SSD-1J} & {\it OSD-1J} & {\it 3L-0J} & {\it 3L-1J} & {\it 4L} \\
\hline \hline
$n_{\ell}$ & $=1$ & $=1$ & $=2$ & $=2$ & $=3$ & $=3$ & $=4$ \\
\hline
$|\sum Q_\ell|$ & $=1$ & $=1$ & $=2$ & $=0$ & -- & -- & -- \\
\hline
$n_{J_{W/Z/h}}$ & $=1$ & $=2$ & $\geq 1$ & $\geq 1$ & $=0$ & $=1$ & $\geq 0$ \\
\hline
$p_T(\ell_0)$ & $>$ 200 & $>$ 200 & $>$ 400 & $>$ 400 & $>$ 300 & $>$ 300 & $>$ 400 \\
\hline
$p_T(\ell_1)$ & -- & -- & $>$ 100 & $>$ 100 & $>$ 100 & $>$ 100 & $>$ 200 \\
\hline
$p_T^{\rm miss}$ & $>$ 100 & $>$ 100 & -- & -- & $>$ 100 & -- & -- \\
\hline
Low $m_{\ell \ell}^{\rm OSSF}$ veto & -- & -- & $>12$ & $>12$ & $>12$ & $>12$ & $>12$ \\
\hline
$Z$-veto & -- & -- & Yes & Yes & -- & -- & -- \\
\hline \hline 
\end{tabular} 
}
\caption{\label{preselection} The preselection criteria for the considered analysis channels. All the selection cuts are in GeV. The symbol ``--" means no requirement is made.}
\end{table}

\begin{figure}[htb!]
\centering
\includegraphics[width=0.49\columnwidth]{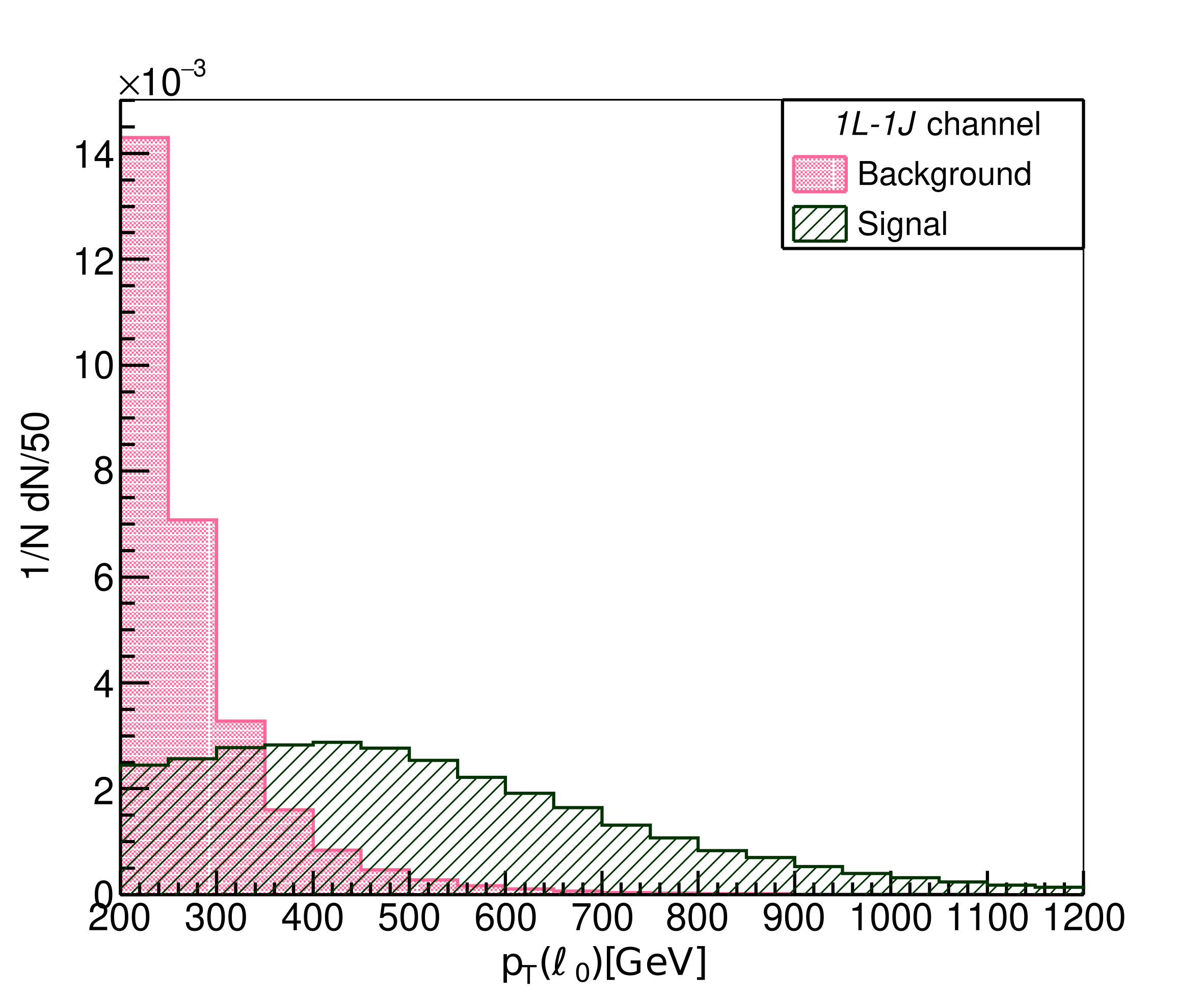}
\includegraphics[width=0.49\columnwidth]{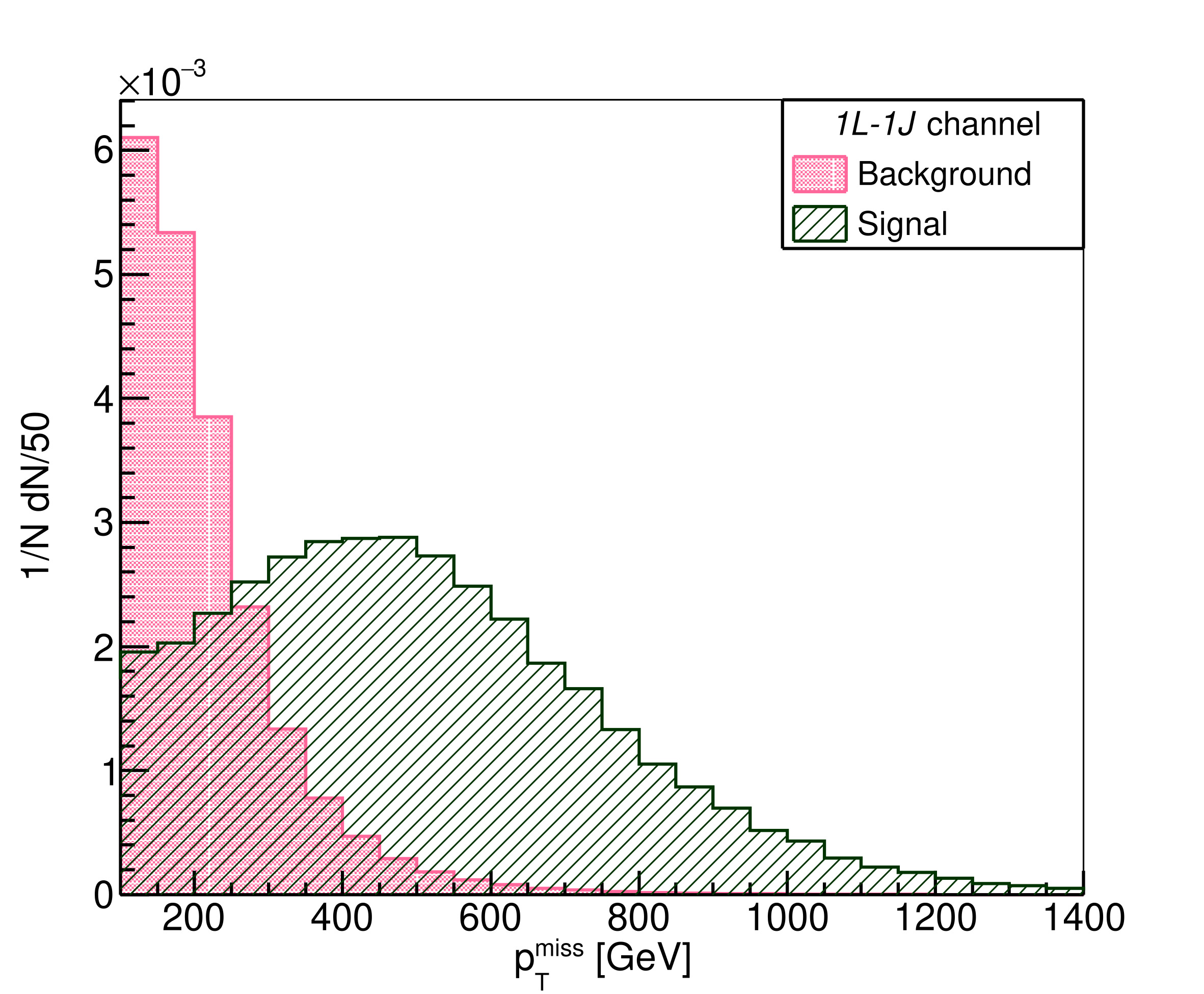}
\includegraphics[width=0.49\columnwidth]{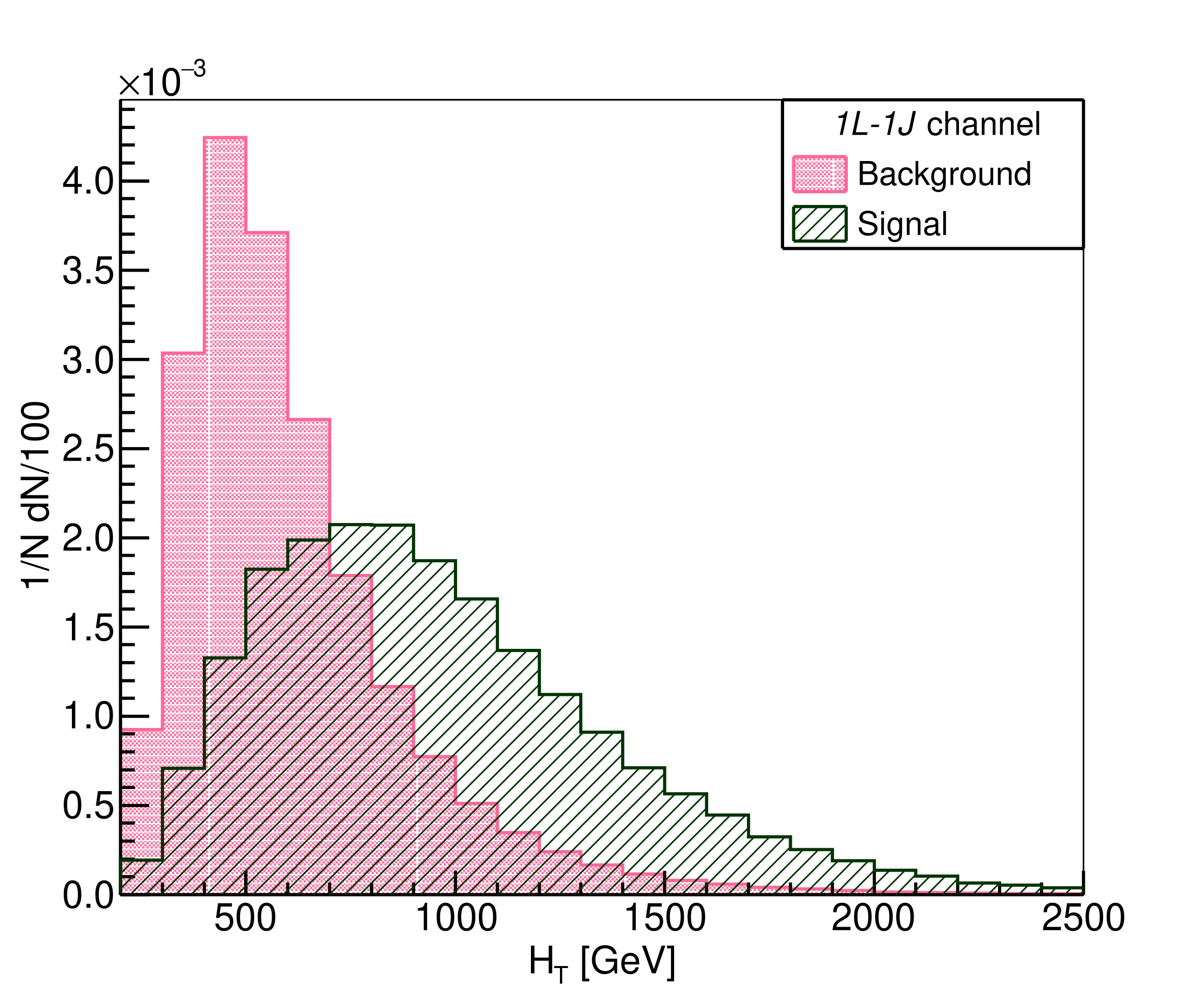}
\includegraphics[width=0.49\columnwidth]{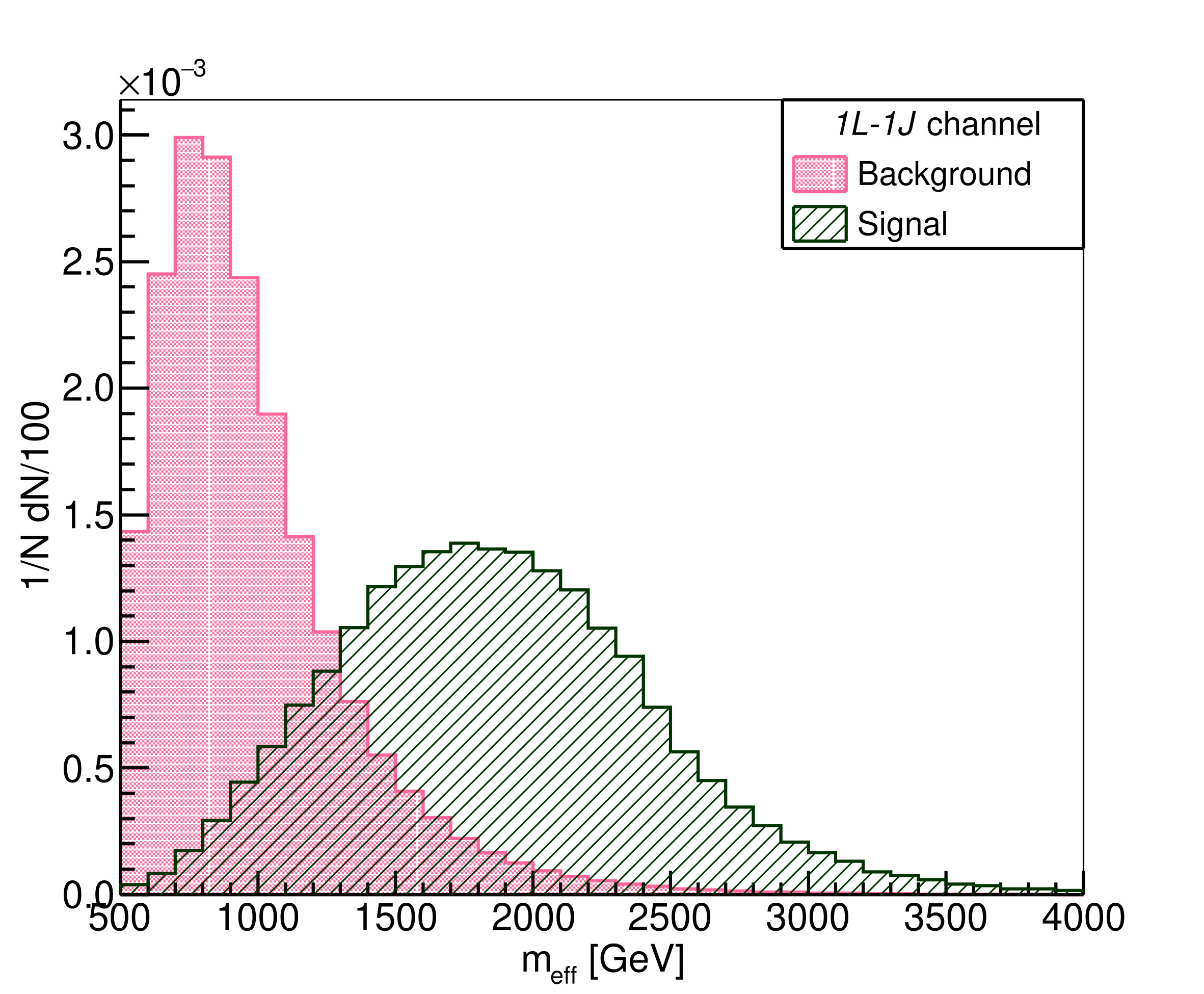}
\includegraphics[width=0.49\columnwidth]{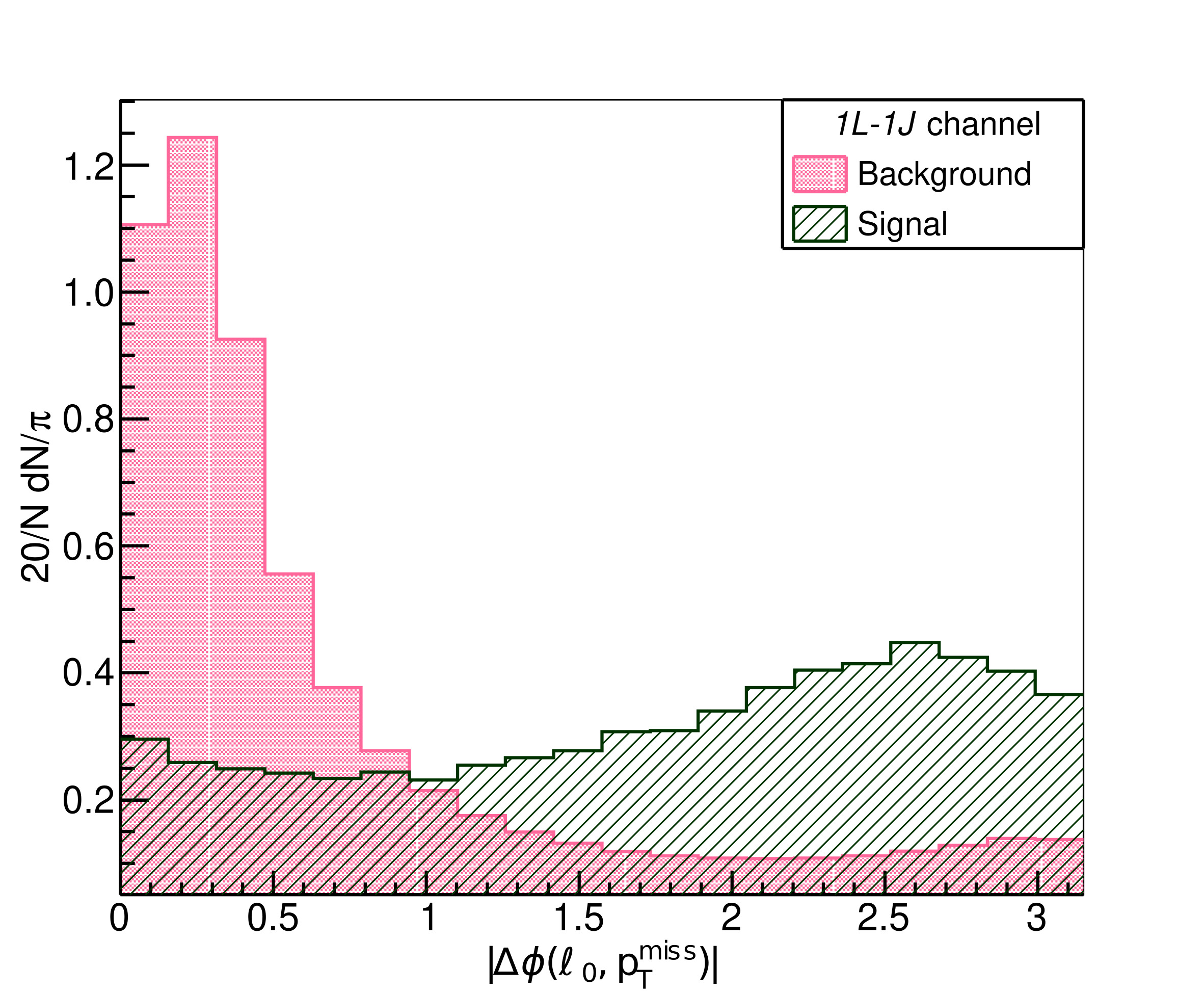}
\includegraphics[width=0.49\columnwidth]{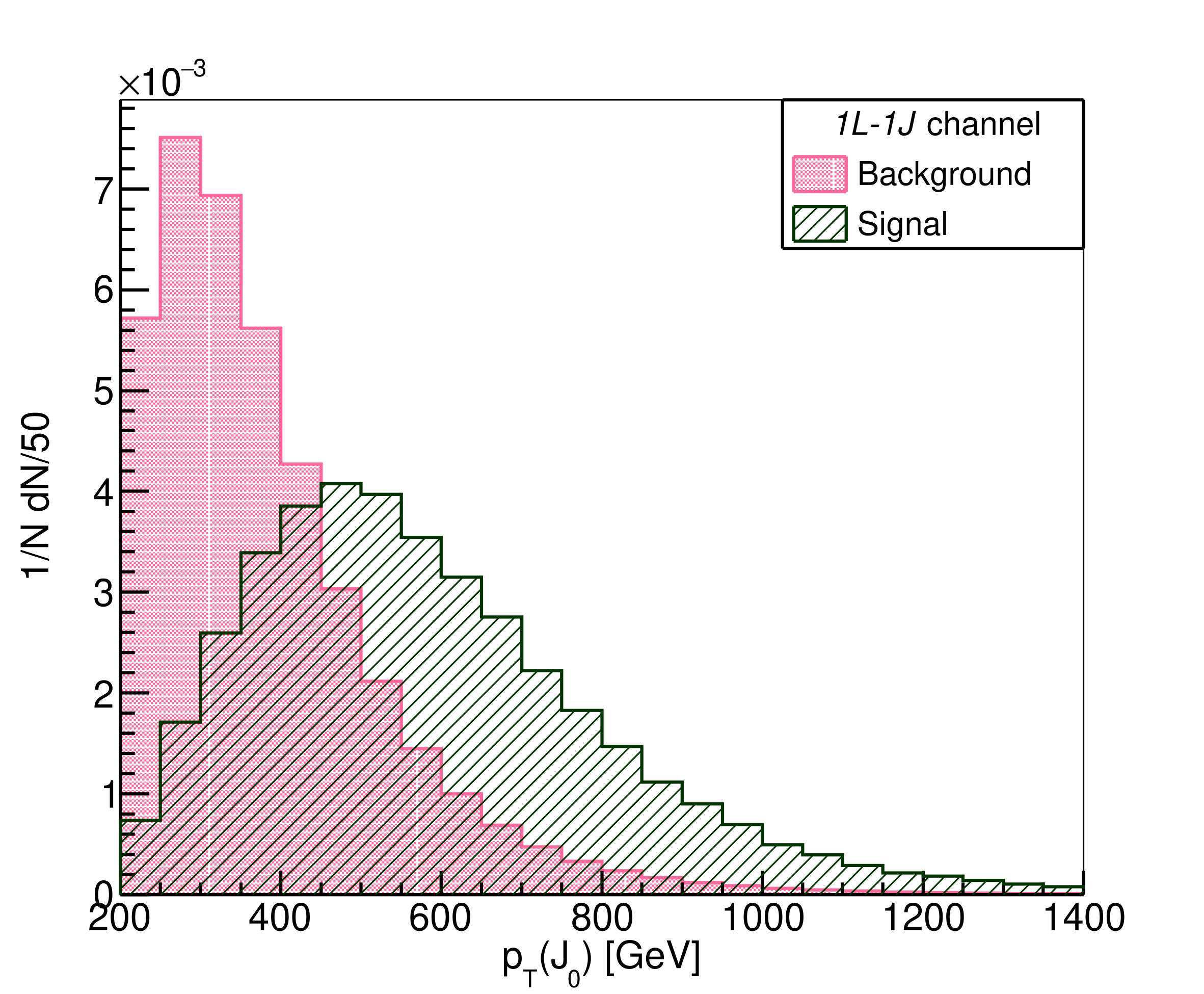}
\caption{Normalised kinematic distributions for signal ($m_\Sigma=1150$ GeV) and background in the {\it 1L-1J} channel. Top panel: lepton $p_T$ (left) and $p_T^{\rm miss}$ (right); middle panel: $H_T$ (left) and $m_{\rm eff}$ (right); bottom panel: $\Delta \phi (\ell_0,p_T^{\rm miss})$ (left) and $W/Z/h$-fat-jet $p_T$ (right).}
\label{fig:1L1J}
\end{figure}

We next briefly discuss the channel-specific selection criteria, which appreciably reduce the remaining SM backgrounds without impinging much on the signal strength. We use various kinematic distributions as guiding premises to choose the appropriate selection cuts.

We display different normalised kinematic distributions for the {\it 1L-1J} events in Figure~\ref{fig:1L1J} for a benchmark point: $m_\Sigma=1150$ GeV. In the top panel, we show the $p_T(\ell_0)$ (left) and $p_T^{\rm miss}$ (right) distributions. The middle panel displays the $H_T$ (left) and $m_{\rm eff}=L_T+H_T+p_T^{\rm miss}$ (right) distributions, where $L_T(H_T)$ is the scalar sum of all the leptons' (jets') $p_T$. In the bottom panel, the left plot shows the distribution of the azimuthal separation between $\ell_0$ and $p_T^{\rm miss}$, $|\Delta \phi (\ell_0, p_T^{\rm miss})|$, whereas the right plot shows the $p_T(J_0)$ distribution. %For an inordinately large fraction of the signal events, $\ell_0$ and $p_T^{\rm miss}$ are largely separated in the azimuth plane, which is not the case for the background events. 
As expected, the kinematic distributions for the signal are much harder than those for the background, see Figure~\ref{fig:1L1J}. Consequently, harder cuts on these kinematic variables are expected to suppress the remaining background without impinging much on the signal strength. On the other hand, $|\Delta \phi (\ell_0, p_T^{\rm miss})|$ turns out to be efficient in discriminating the background (in particular, the background from single W-boson) and signal. We impose the following selection cuts:
\begin{align*}
\textbf{\textit{SI-1}:} ~ & p_T(\ell_0) > 400 {\rm ~GeV~and~} p_T^{\rm miss} > 400 {\rm ~GeV},
\\
\textbf{\textit{SI-2}:} ~ & H_T > 500 {\rm ~GeV,~} m_{\rm eff} > 1800 {\rm ~GeV~and~} 
\\
& p_T(J_0) > 300 {\rm ~GeV},
\\
\textbf{\textit{SI-3}:} ~ & |\Delta \phi (\ell_0, p_T^{\rm miss})| > 0.5~.
\end{align*}

\begin{figure}[htb!]
\centering
\includegraphics[width=0.75\columnwidth]{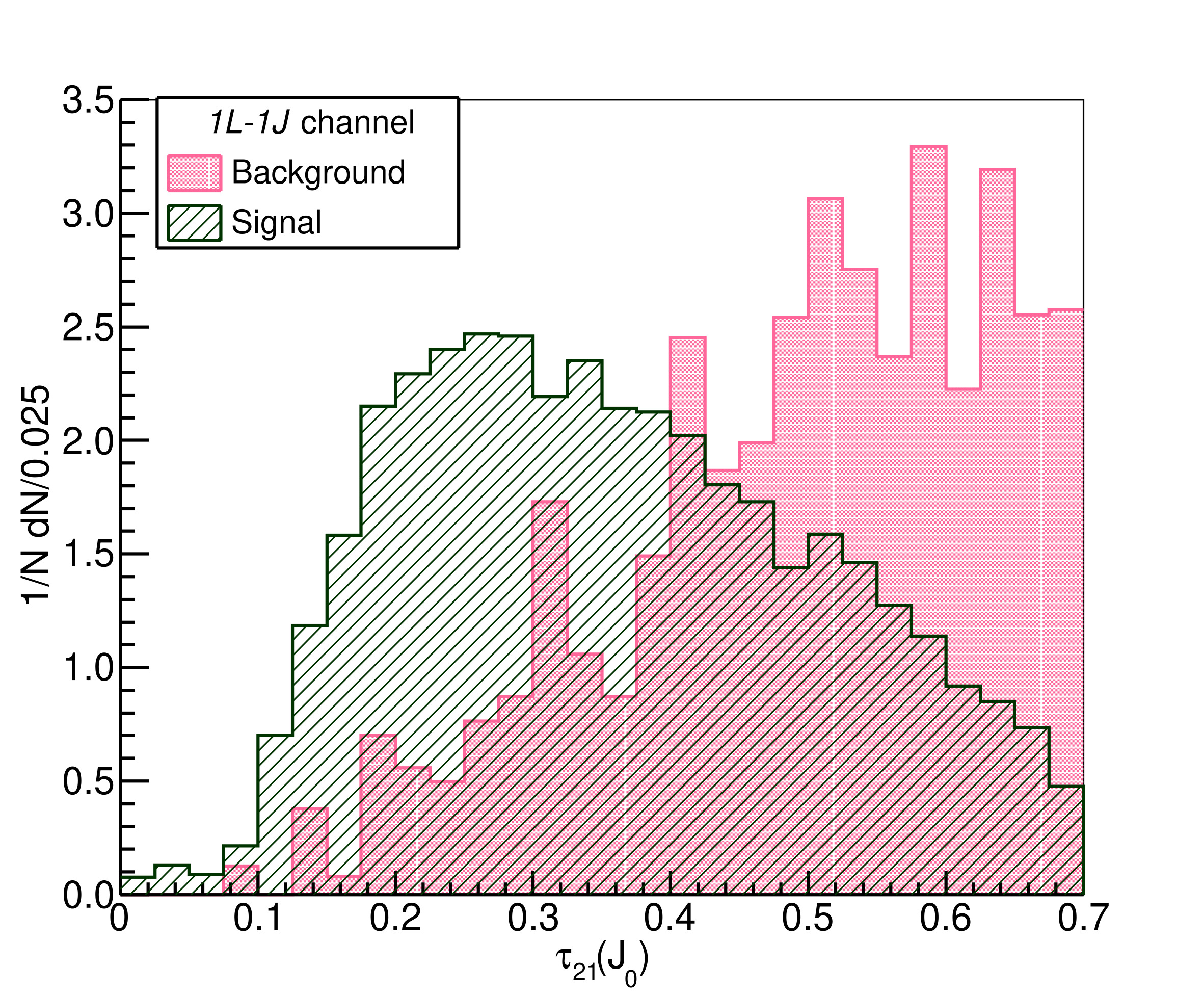}
\caption{Normalised distribution of $\tau_{21}(J_0)$ after passing the {\it S1-1}, {\it S1-2} and {\it S1-3} cuts for signal ($m_\Sigma=1150$ GeV) and background in the {\it 1L-1J} channel.}
\label{fig:1L1Jb}
\end{figure}

\noindent Also displayed, in Figure~\ref{fig:1L1Jb}, is the normalised distribution of $\tau_{21}(J_0)$ for the {\it 1L-1J} events passing the {\it S1-1}, {\it S1-2} and {\it S1-3} cuts. As expected, an inordinately large fraction of the signal events have lower $\tau_{21}$ as compared to the remaining background events. Therefore, to further improve the signal-to-background ratio, we impose the following selection cut:
\begin{align*}
\textbf{\textit{SI-4}:} ~ & \tau_{21}(J_0) < 0.5~.
\end{align*}

We display the effect of the selection cuts on the signal and background strengths for the {\it 1L-1J} channel in Table~\ref{table:cutflow}. It shows the progression of the number of expected background and signal ($m_\Sigma=1150$ GeV) events as the subsequent selection cuts are imposed for 1000 fb$^{-1}$ luminosity at the 13 TeV LHC. As we expected, the preselection cuts and channel-specific selection cuts effectively vanquish the SM backgrounds while keeping a large fractions of the signal.

\begin{table}[htb!]
\centering
\scalebox{0.91}{
\begin{tabular}{|l|l|*{4}{p{1cm}|}}
\hline
\hline
\multicolumn{6}{|l|}{\textbf{\textit{1L-1J} channel}} \\
\hline
\hline
Event sample & {\it Preselection} & {\it SI-1} & {\it SI-2} & {\it SI-3} & {\it SI-4} \\
\hline
Drell-Yan & $\sim 2.9 \times 10^5$ & 5119 & 4057 & 221 & 98 \\
\hline
$t\bar{t}$ & $\sim 1.7 \times 10^5$ & 291 & 190 & 44 & 20 \\
\hline
Other backgrounds & $\sim 1.7 \times 10^4$ & 239 & 185 & 25 & 14 \\
\hline
Total background & $\sim 4.8 \times 10^5$ & 5649 & 4432 & 290 & 132 \\
\hline
Signal & 330 & 126 & 93 & 90 & 71 \\
\cline{1-6} \noalign{\vskip\doublerulesep \vskip-\arrayrulewidth} \cline{1-6}
\end{tabular} 
}
\caption{\label{table:cutflow} The number of expected background and signal ($m_\Sigma=1150$ GeV) events in the {\it 1L-1J} channel after passing various cuts for 1000 fb$^{-1}$ luminosity at the 13 TeV LHC.}
\end{table}

After discussing the cuts for the {\it 1L-1J} channel in detail, we next summarise the channel specific selection cuts for all other channels in Table~\ref{table:cuts}. Table~\ref{table:cuts} also shows the number of expected background and signal ($m_\Sigma=1150$ GeV) events after all the selection cuts are imposed for 1000 fb$^{-1}$ luminosity at the 13 TeV LHC.

\begin{table}[htb!]
\centering
\scalebox{0.87}{
\begin{tabular}{|l|p{5.5cm}|l|l|}
\hline
\hline
Channel & Selection cuts & Background & Signal \\
\hline
\textit{1L-2J} & $p_T(\ell_0) > 400$, $p_T^{\rm miss} > 300$, & 74 & 63\\
 & $|\Delta \phi (\ell_0, p_T^{\rm miss})| > 0.5$ & & \\
\hline
\textit{SSD-1J} & $L_T > 600$, $H_T > 500$, $m_{\rm eff} > 1600$, & 81 & 32 \\
 & $p_T(J_0) > 300$ & & \\
\hline
\hline
\textit{OSD-1J} & $L_T > 700$, $H_T > 500$, $m_{\rm eff} > 2000$, & 131 & 46\\
 & $p_T(J_0) > 400$, $\tau_{21}(J_0) < 0.6$ & & \\
\hline
\hline
\textit{3L-0J} & $L_T > 600$, $m_{\rm eff} > 1500$, $L_T/H_T > 1.0$, & 67 & 24\\
 & $p_T^{\rm miss}/H_T > 0.4$ & & \\
\hline
\hline
\textit{3L-1J} & $L_T > 700$, $m_{\rm eff} > 1600$, $p_T(J_0) > 300$, & 28 & 20\\
& $L_T/H_T > 1.0$ & & \\
\hline
\hline
\textit{4L} & $L_T > 1000$, $m_{\rm eff} > 1600$ & 9 & 8\\
\hline
\hline
\end{tabular} 
}
\caption{\label{table:cuts} Summary of the channel-specific selection cuts along with the number of expected background and signal ($m_\Sigma=1150$ GeV) events after all the selection cuts are imposed for 1000 fb$^{-1}$ luminosity at the 13 TeV LHC. All the dimensionful cuts are in GeV.}
\end{table}

In order to enhance the sensitivity of this search, the selected events in each channel are divided into several independent search bins using a primary kinematic discriminant between the signal and background. The discriminants used in different channels are as follows:
\begin{itemize}
\item \textbf{\textit{1L-1J:}} the invariant mass of the lepton and the fat-jet: $m_{\ell J}=m_{\ell_0 J_0}$;
\item \textbf{\textit{1L-2J:}} the invariant mass of the lepton and the farthest fat-jet in the azimuth plane: 
\begin{align*}
m_{\ell J} &= m_{\ell_0 J_0} {\rm ~if~} \Delta \phi (\ell_0, J_0) > \Delta \phi (\ell_0, J_1) \\
           &= m_{\ell_0 J_1} {\rm ~otherwise};
\end{align*}
\item \textbf{\textit{SSD-1J}} and \textbf{\textit{OSD-1J:}} $(i)$ for events with exactly one $J_{W/Z/h}$: the invariant mass of the fat-jet and the farthest lepton in the azimuth plane:
\begin{align*}
m_{\ell J} &= m_{\ell_0 J_0} {\rm ~if~} \Delta \phi (\ell_0, J_0) > \Delta \phi (\ell_1, J_0) \\
           &= m_{\ell_1 J_0} {\rm ~otherwise};
\end{align*}
$(ii)$ for events with more than one $J_{W/Z/h}$: 
\begin{align*}
\quad \qquad m_{\ell J} &= m_{\ell_0 J_0} {\rm ~if~} |m_{\ell_0 J_0}-m_{\ell_1 J_1}| < |m_{\ell_0 J_1}-m_{\ell_1 J_0}| \\
           &= m_{\ell_0 J_1} {\rm ~otherwise};
\end{align*}
\item \textbf{\textit{3L-0J}}: Missing neutrinos do not allow kinematic reconstruction of the triplets in this channel, and, thus, we use $L_T+p_T^{\rm miss}$ as primary kinematic discriminant;
\item \textbf{\textit{3L-1J}}: $(i)$ for events with one on-$Z$ lepton pair, the invariant mass of the trilepton system: $m_{\ell_0 \ell_1 \ell_2}$;\\
$(ii)$ for events with no on-$Z$ lepton pair: the invariant mass of the fat-jet and the farthest lepton in the azimuth plane;
\item \textbf{\textit{4L}}: the invariant mass of the on-$Z$ lepton pair and the farthest lepton in the azimuth plane: $m_{(\ell \ell)_Z \ell}$.
\end{itemize}

%As discriminant, we use the invariant mass of the lepton and the fat-jet ($m_{\ell_0 J_0}$) for the {\it 1L-1J} channel, while for the {\it 1L-2J} channel, the invariant mass of the lepton and the farthest fat-jet in the azimuth plane ($m_{\ell_0 J_0}$ or $m_{\ell_0 J_1}$) is used. For the {\it SSD-1J} and {\it OSD-1J} events with exactly one $J_{W/Z/h}$, the invariant mass of the fat-jet and the farthest lepton in the azimuth plane ($m_{\ell_0 J_0}$ or $m_{\ell_1 J_0}$) is used; while for those with two $J_{W/Z/h}$, we use $m_{\ell_0 J_0}$ if $|m_{\ell_0 J_0}-m_{\ell_1 J_1}| < |m_{\ell_0 J_1}-m_{\ell_1 J_0}|$ or $m_{\ell_0 J_1}$ if otherwise.
 
\noindent For all the channels except the {\it 3L-0J} one, the distribution of the primary discriminant peaks at the triplet mass and thus could be used to reconstruct the same. We summarise the binning scheme in Table~\ref{table:binning}, which yields a total of 50 statistically independent search bins. The width of the bins is chosen to ensure smooth behaviour for the expected background and signal distributions. The underflow (overflow) events are contained in the first (last) bin for each channel. 

\begin{table}[htb!]
\centering
\scalebox{0.9}{
\begin{tabular}{|l c c c|}
\hline
Channel & Kinematic discriminant & Range (GeV) &  Number of bins 
\\
\hline
{\it 1L-1J} & $m_{\ell_0 J_0}$ & $[600,2000]$ & 7
\\
{\it 1L-2J} & $m_{\ell_0 J_0}$ or $m_{\ell_0 J_1}$ & $[600,2000]$ & 7
\\
{\it SSD-1J} & $m_{\ell_0 J_0}$ or $m_{\ell_1 J_0}$ & $[600,2000]$ & 7
\\
{\it OSD-1J} & $m_{\ell_0 J_0}$ or $m_{\ell_1 J_0}$ & $[600,2000]$ & 7
\\
{\it 3L-0J} & $L_T+p_T^{\rm miss}$ & $[750,2750]$ & 8
\\
{\it 3L-1J} & $m_{\ell_0 J_0}$ or $m_{\ell_1 J_0}$ or $m_{\ell_0 \ell_1 \ell_2}$ & $[600,2000]$ & 7
\\
{\it 4L} & $m_{(\ell \ell)_Z \ell}$ & $[600,2000]$ & 7
\\
\hline
\end{tabular}
}
\caption{Binning scheme of different channels.}
\label{table:binning}
\end{table}

\begin{figure}[htb!]
\centering
\includegraphics[width=0.49\columnwidth]{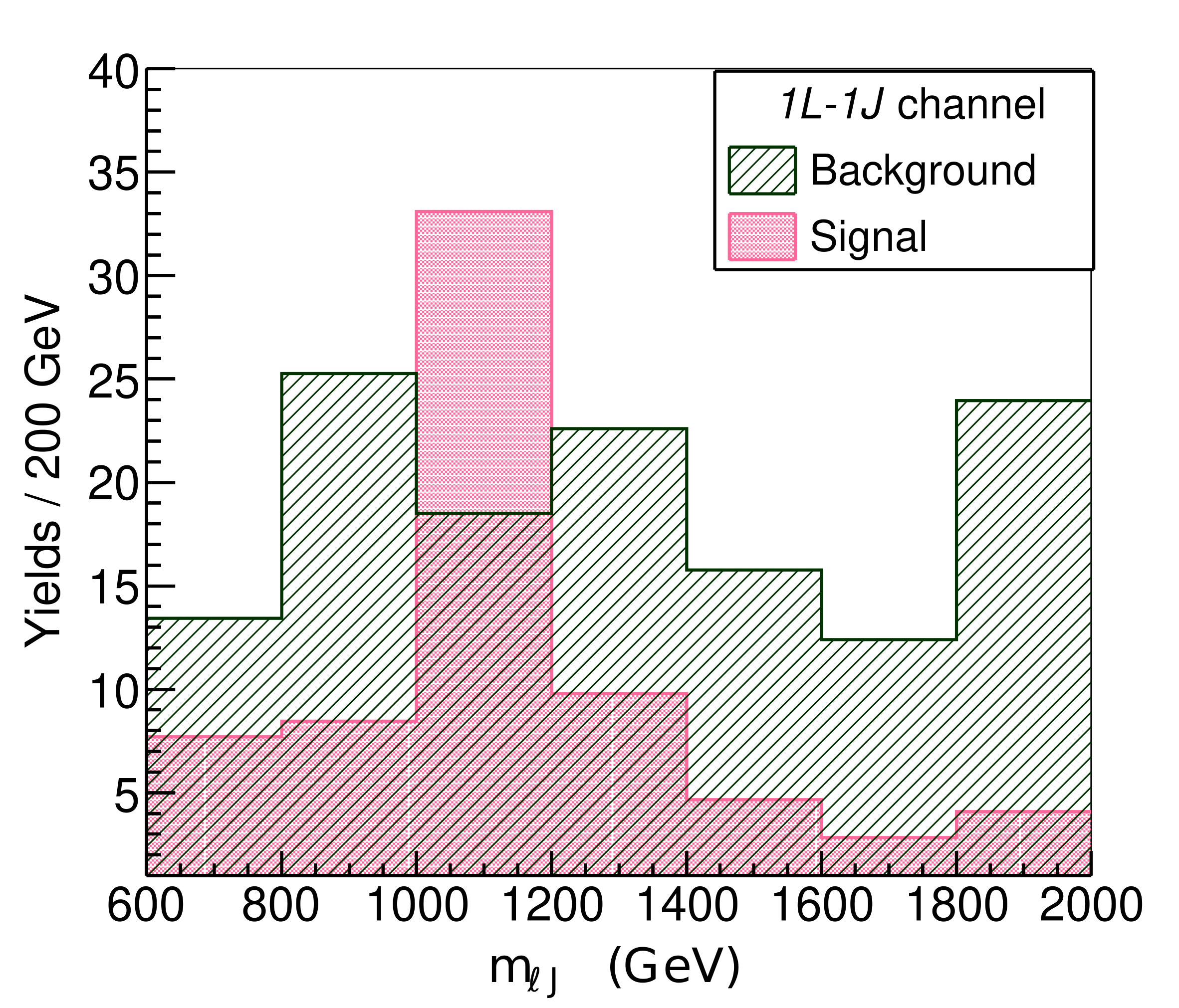}
\includegraphics[width=0.49\columnwidth]{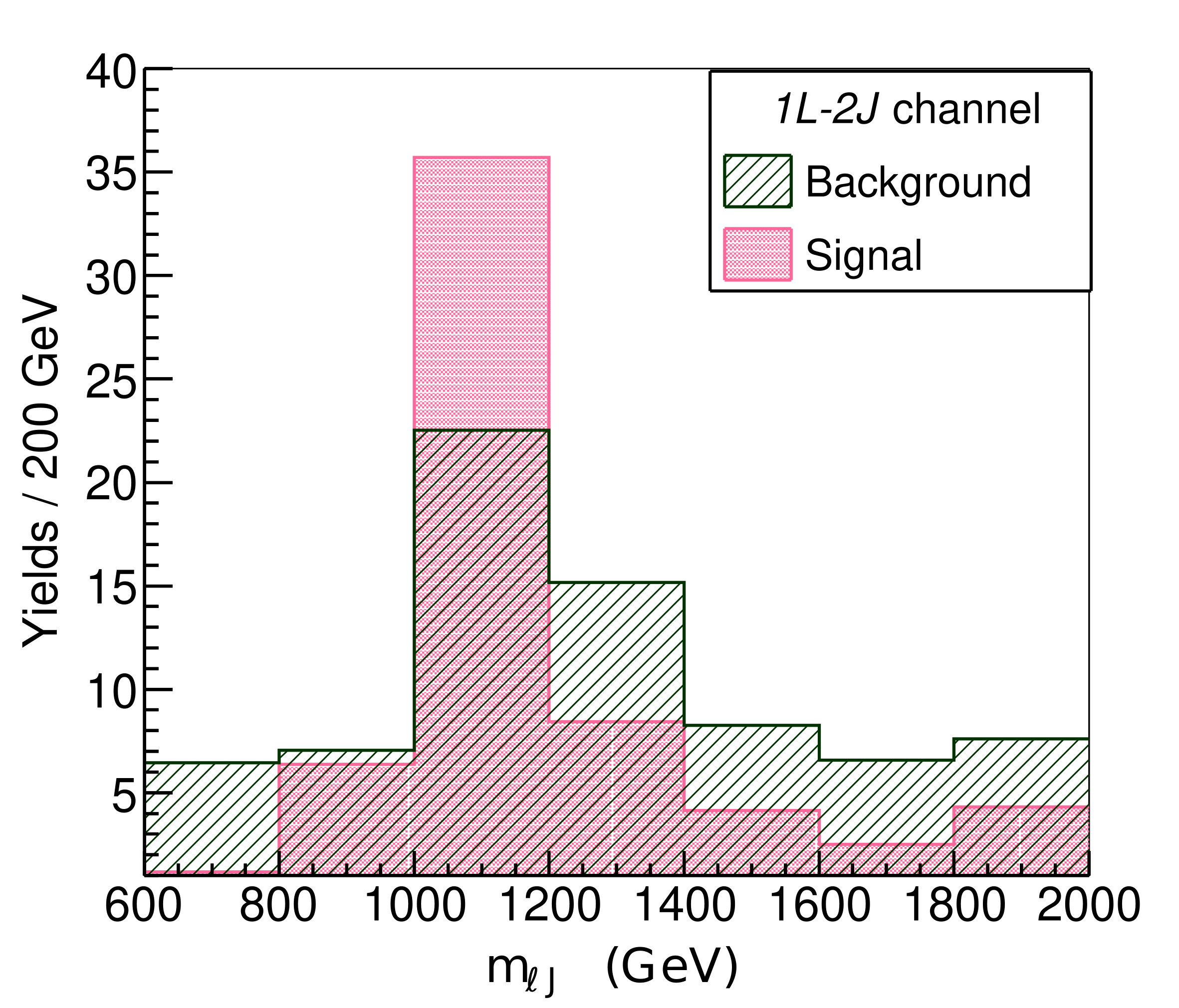}
\includegraphics[width=0.49\columnwidth]{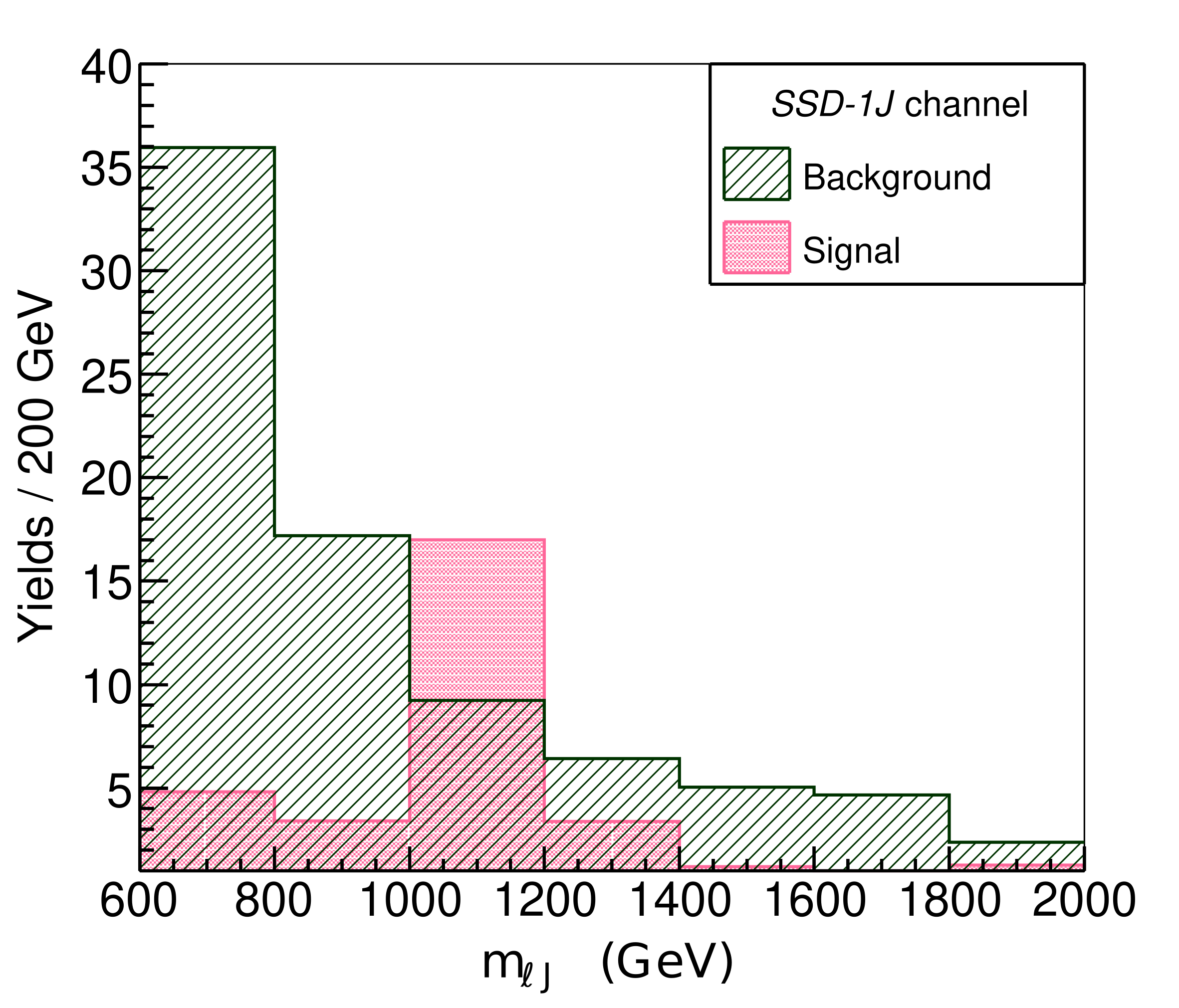}
\includegraphics[width=0.49\columnwidth]{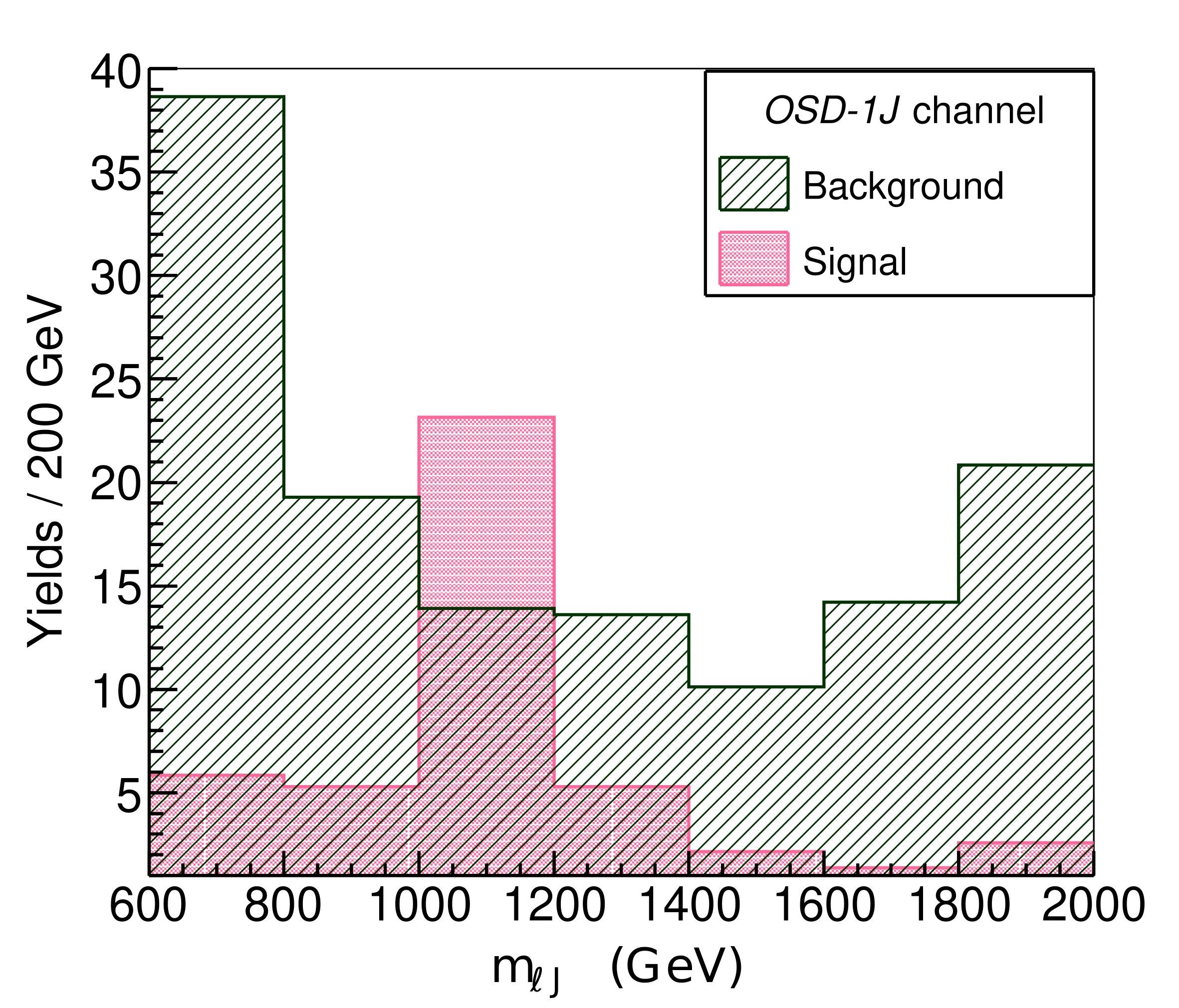}
\includegraphics[width=0.49\columnwidth]{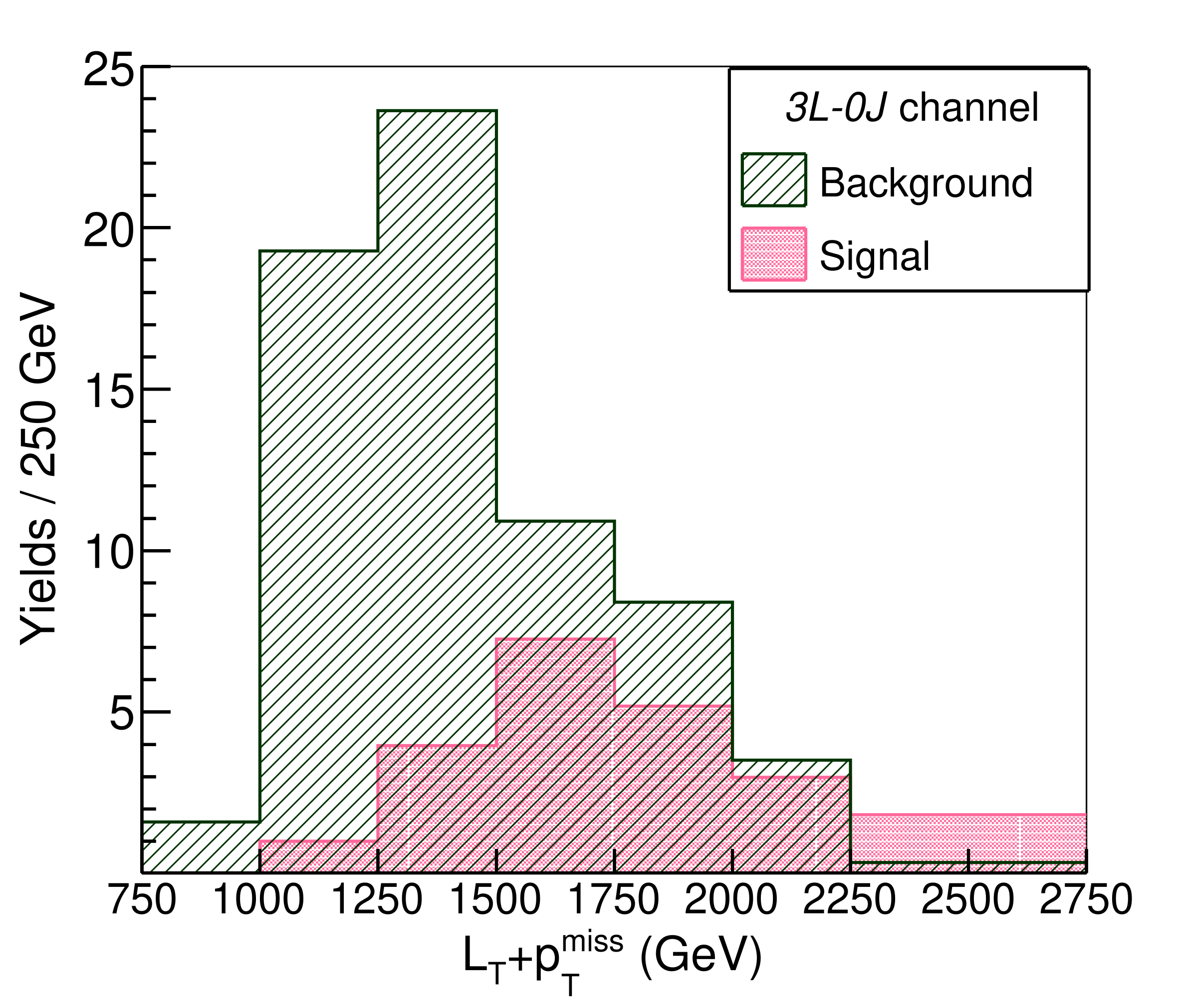}
\includegraphics[width=0.49\columnwidth]{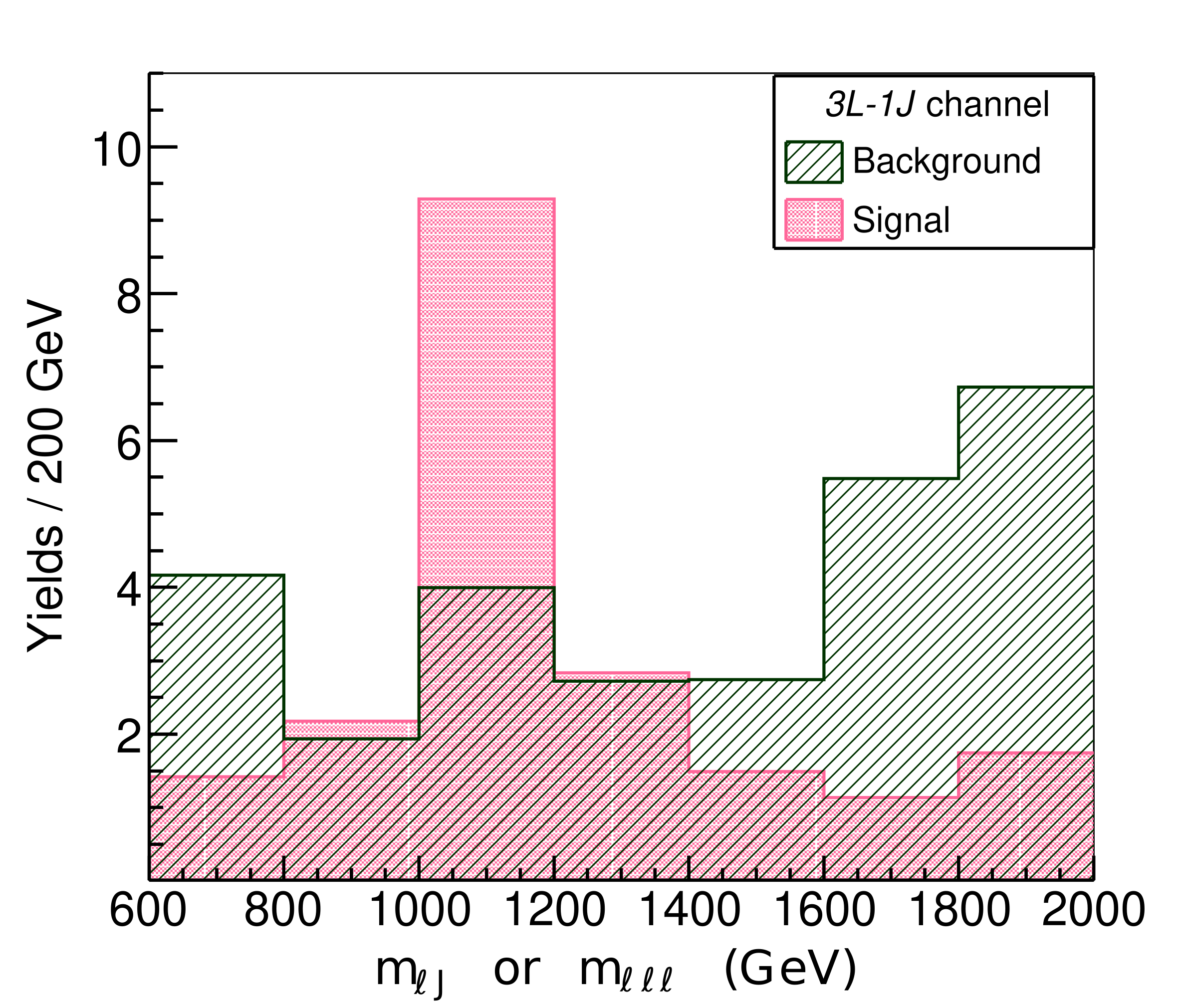}
\includegraphics[width=0.49\columnwidth]{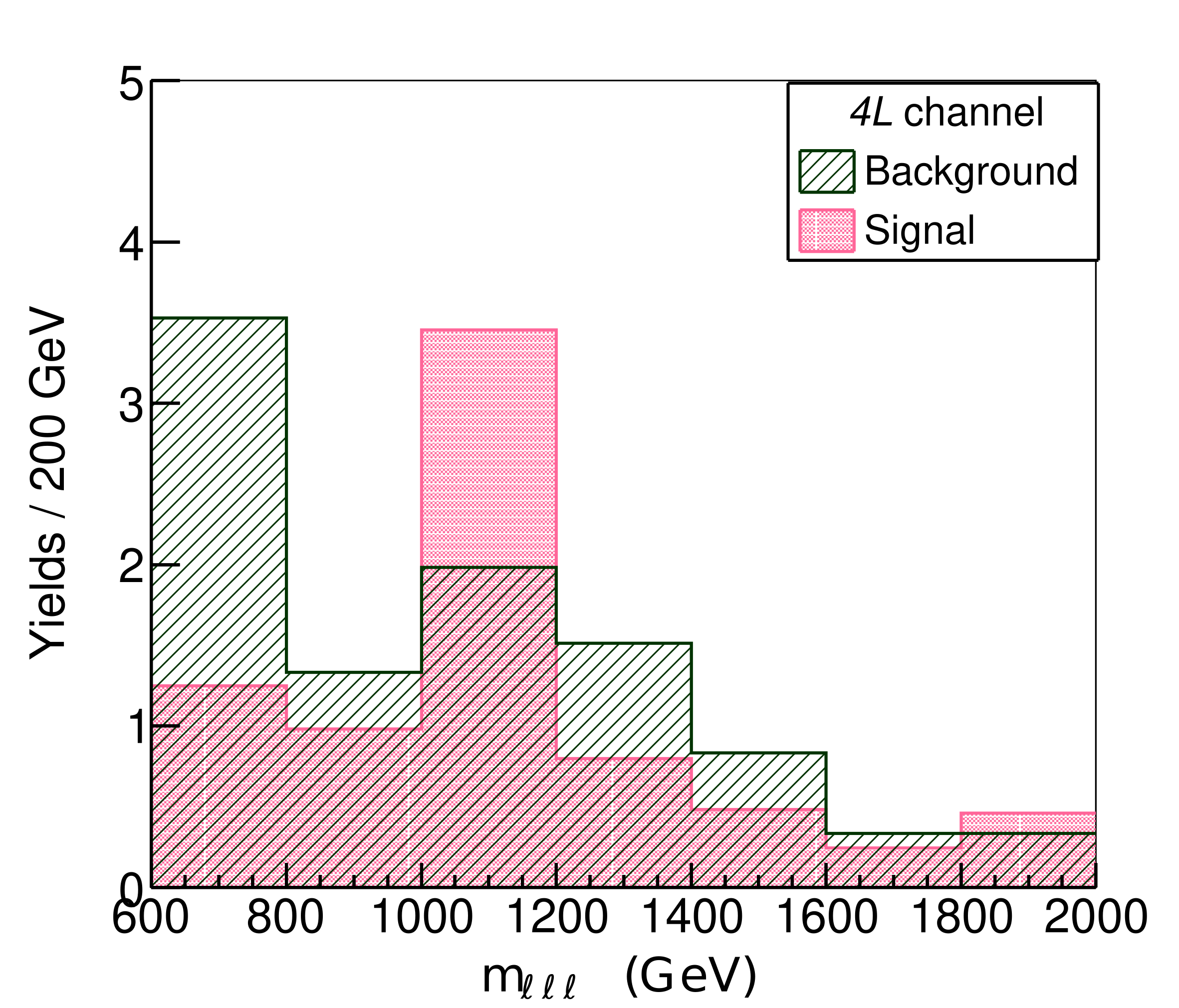}
\caption{Distributions of the primary kinematic discriminants for for the signal ($m_\Sigma = 1150$ GeV) and background. The events are weighted for 1000 fb$^{-1}$ luminosity at the 13 TeV LHC.}
\label{fig:SRs}
\end{figure}

Figure~\ref{fig:SRs} displays the distributions of the primary kinematic discriminants for the signal ($m_\Sigma = 1150$ GeV) and background events for all the channels. The events are weighted for 1000 fb$^{-1}$ luminosity at the 13 TeV LHC. As expected, these distributions (except the one for {\it 3L-0J} channel) peak in 1000--1200 GeV bin, and thereby reconstructing the triplet mass; a simple rebinning of these distributions with smaller bin widths would make this mass reconstruction apparent.

\subsection{Discovery reach}
\begin{figure}[htb!]
\centering
\includegraphics[width=0.9\columnwidth]{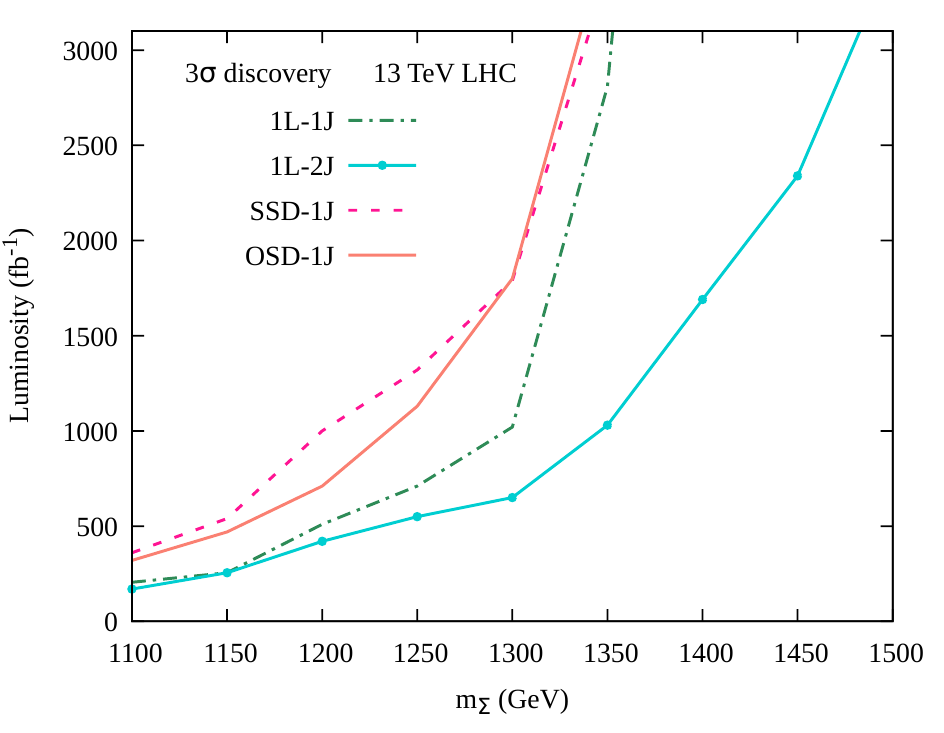}
\caption{Required Luminosity for $3\sigma$ discovery of the triplet fermions in different analysis channels.}
\label{fig:lumi2}
\end{figure}

We next estimate the discovery reach of the present search for the triplet fermions. To this end, we use a hypothesis tester named {\it Profile Likelihood Number Counting Combination} which uses the library {\it RooFit} \cite{Verkerke:2003ir} in the {\it ROOT} \cite{Brun:1997pa} environment. This hypothesis tester considers all the search bins as independent number counting channels with uncorrelated systematic errors. The uncertainties are included via the {\it Profile Likelihood Ratio}. For the sake of simplicity, we assume an overall 20\% total uncertainty on the estimated background. Further, to avoid spurious exclusions/discoveries and to ensure robustness in statistical interpretations, we replace the less than one per-bin expected background yield at 3000 fb$^{-1}$ with one background yield. This renders our estimated significances to be a bit conservative. In Figure~\ref{fig:lumi2}, we project the required luminosities for $3\sigma$ discovery of the triplet fermions in different analysis channels as a function of their mass; for brevity, we display the same for the most promising channels only. We see that the {\it 1L-2J} channel is the most promising one with the discovery reach of 1480 GeV, while the rest of the channels have a similar discovery reach of 1330-1350 GeV.\footnote{Note that the {\it 1L}-channels are not suppressed by the leptonic branching fractions of the SM bosons, and thus, yield a sufficient number of signal events even for larger triplet mass. Moreover, not only does the requirement of two $J_{W/Z/h}$ in the final state vanquish the background, but it also allows kinematic reconstruction of the triplets without any ambiguity (see the top panel in Figure~\ref{fig:SRs}). On the contrary, final states with higher lepton multiplicity (three or more) being suppressed by the leptonic branching fractions of the SM bosons are less sensitive in probing triplets with higher masses. Note that the recent CMS/ATLAS searches targetting the triplet fermions in the type-III see-saw model rely on multi-lepton final states only, and thus, are not sensitive enough in probing them with higher masses.}

\begin{figure}[htb!]
\centering
\includegraphics[width=0.9\columnwidth]{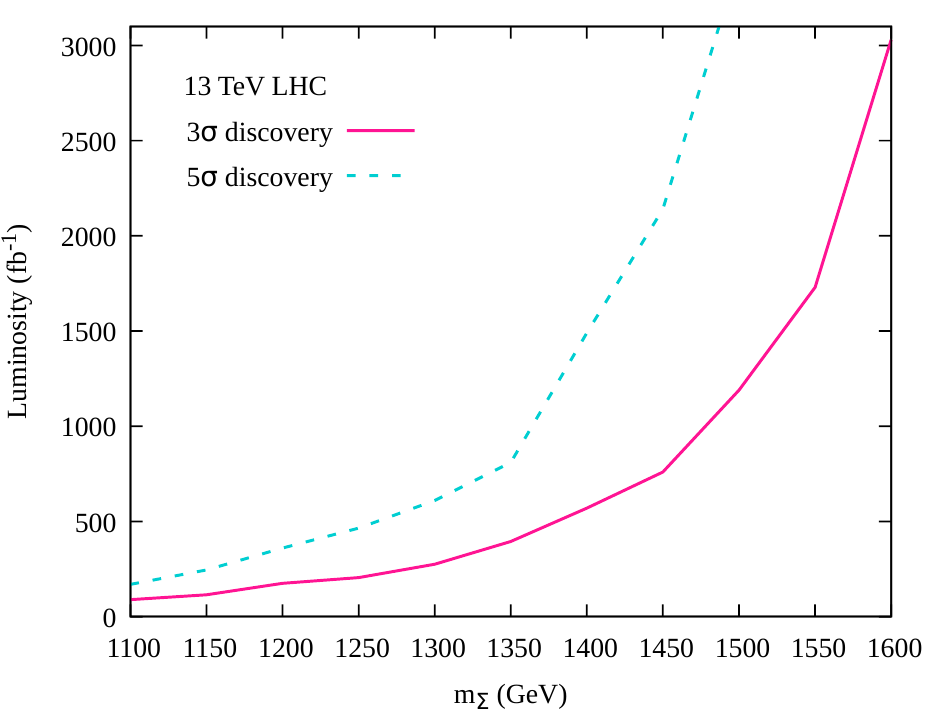}
\caption{Required Luminosity for $3\sigma$ and $5\sigma$ discovery of the triplet fermions in the combined channels.}
\label{fig:lumi}
\end{figure}

Since all the analysis channels considered in this search are mutually exclusive, it is reasonable to combine them. For the combined channels, we project the required luminosities for both $3\sigma$ and $5\sigma$ discoveries of the triplet fermions as a function of their mass in Figure~\ref{fig:lumi}. The $5\sigma$ ($3\sigma$) discovery reach of this search is about 1265 (1380) and 1480 (1600) GeV, respectively, at 500 and 3000 fb$^{-1}$ luminosities.

\section{\label{sec:conclusion}Summary}
The CMS and ATLAS collaborations have performed several searches targetting the triplet fermions in the type-III see-saw model. However, the final states considered in these searches being beset with considerably large SM backgrounds are deemed not sensitive enough to probe the triplet fermions much heavier than 1 TeV. To this end, we perform a search for the triplet fermions in final states with multiple leptons and fat-jets that are cleaner than the usual LHC searches and allow reconstruction of the triplet mass. After performing a systematic and comprehensive search with seven distinct analysis channels, we project the required luminosities for both $3\sigma$ and $5\sigma$ discoveries of the triplet fermions as a function of their mass. The triplet fermions with mass as large as 1265 (1380) and 1480 (1600) GeV could be discovered with $5\sigma$ ($3\sigma$) significance at 500 and 3000 fb$^{-1}$ luminosities, respectively. \footnote{Thogh single production of the triplet fermions, in principle, in association with a SM lepton is possible, such processes are extremely suppressed both at $pp$ and $e^-e^+$ colliders on account of their small mixing with the SM leptons. In view of pair production of the triplet fermions, in the most optimistic scenario, a high energy $e^-e^+$ collider could probe them very close to its kinematic reach \cite{Goswami:2017jqs,Das:2020gnt}, {\it i.e.} 1500 GeV for a 3 TeV collider. Note that the $3\sigma$ ($5\sigma$) reach of our search at 3000 fb$^{-1}$ luminosity extends beyond (very close to) the kinematic reach of a 3 TeV $e^-e^+$ collider.} In closing, we mention that a search similar to the one presented in this work is prophesied to carry through for a large class of neutrino mass models containing new weak gauge multiplets. We undertake a detailed analysis for the same in a future work \cite{Ashanujjaman:toappear}

\acknowledgments 
KG acknowledges support from the DST INSPIRE Research Grant [DST/INSPIRE/04/2014/002158] and SERB Core Research Grant [CRG/2019/006831]. The simulations were supported in part by the SAMKHYA: High Performance Computing Facility provided by Institute of Physics, Bhubaneswar.

\end{document}